\documentstyle[preprint,aps,eqsecnum]{revtex}
\begin{document}
\def\today{\space\number\day\ \ifcase\month\or
January\or February\or March\or April\or May\or June\or July\or
August\or September\or October\or November\or December\fi
\ \number\year}
\overfullrule=0pt  
\def\mynote#1{{[{\it NOTE: #1}]}}
\def\fEQN#1#2{$$\hbox{\it #1\hfil}\EQN{#2}$$}
\def\Acknowledgements{{\bigskip\leftline
{{\bf Acknowledgments}}\medskip}}

\begin{titlepage}



 \vspace{0cm}
 
\begin{center}
{\large\bf  Effective Scalar Field Theory\\ 
and Reduction of Couplings.  }

 \vspace{0.4cm}
 
{\bf Mario Atance}\footnote{E-mail: atance@posta.unizar.es}  and
{\bf Jos\'e~Luis~Cort\'es}\footnote{E-mail: cortes@posta.unizar.es}  

 \vspace{0.2cm}

{\sl Departamento de F\'{\i}sica Te\'orica,\\ 
Universidad de Zaragoza,
50009 Zaragoza, Spain.}

 \vspace{0.2cm}


\end{center}
\vspace{0.2cm}
\begin{abstract}
A general discussion of the renormalization of the quantum theory 
of a scalar field as an effective field theory is presented. The
renormalization group equations in a mass-independent renormalization
scheme allow us to identify the possibility to go beyond the
renormalizable $\phi^{4}$ theory without losing its predictive
power. It is shown that there is a minimal extension with just one 
additional free parameter (the mass scale of the effective theory
expansion) and some of its properties are discussed.

\bigskip\bigskip
\noindent PACS number(s): 11.10.Hi
\bigskip

\end{abstract}

\end{titlepage}
\hfill

\section{Introduction}

Our present understanding of quantum field theory as the low-energy
limit of any relativistic quantum-mechanical system~\cite{Weinberg I}
has changed our point of view on general problems in quantum field
theory such as renormalization~\cite{Polchinski}. An effective field
theory Lagrangian contains an infinite number of terms including the 
usual ones in a renormalizable Lagrangian (in the power-counting 
sense~\cite{Dyson}) corresponding to the terms with dimension less than or 
equal to 4.

A natural way to parametrize the Lagrangian
is based on the introduction of a fixed mass scale $M$, which is a
characteristic scale of the physical system described by the
effective theory, and a dimensionless parameter for each term in the 
Lagrangian giving the corresponding coefficient in units of $M$ raised
to the appropriate power. Ultraviolet divergences can be cancelled by
a renormalization of the infinite number of parameters. If one uses
an appropriate renormalization scheme (a mass-independent
renormalization scheme~\cite{Weinberg II,Brezin,t'Hooft}) then,
when one computes a process at some energy $E$, the parameter associated 
to a term in the Lagrangian of dimension $n+4$ gives a contribution 
proportional\footnote[1]{A very clear discussion on this 
point can be found in~\cite{Manohar1}.} to $(E / M)^{n}$. If terms suppressed
by powers of $(E / M)$ are neglected, the usual renormalizable theory 
result is recovered, and when one computes to a given order in 
$(E / M)$ only a finite number of parameters appear. In this sense,
although the effective theory has an infinite number of parameters,
the theory has predictive power~\cite{EFT}. When the energy
becomes comparable to the mass scale $M$ of the effective theory
one goes beyond the domain of validity of the effective field theory 
expansion and one has to consider a new theory, either a new field
theory incorporating the appropriate fields to describe the degrees
of freedom at these energies, or a theory going beyond the general
principles of quantum mechanics and special relativity.

In order to be able to cancel all ultraviolet divergences one 
usually considers all the terms compatible with some symmetry
principles. But this is not necessarily the case. The 
possibility to have a renormalizable theory with a reduced 
number of parameters (method of reduction of couplings) 
has been studied in recent years for different 
purposes.\footnote[2]{For a recent review with a list of references 
see~\cite{Oehme}.} The program of reduction of couplings was
initiated in~\cite{Oehme-Zimmerman} by looking for massless
renormalizable theories in the power-counting sense with a
single dimensionless coupling parameter. The same idea can
be applied in the case of effective field 
theories\footnote[3]{The idea of considering a reduction of couplings 
in a non renormalizable theory appears for the first time 
in~\cite{Toro}.} looking for relations between the  renormalized couplings
compatible with the renormalization group equations.

In previous works~\cite{AC1,AC2} the authors considered the
possibility to apply the method of reduction of couplings to
the effective field theoretic formulation of quantum gravity.
The nonrenormalizability of the theory is not an obstacle to 
identifying a theory with a finite number of independent 
parameters. In order to get this result one has to assume that 
the mass scale associated to the Newtonian limit (Planck mass) 
is much larger than the mass scale of the effective theory  
and one has to neglect all the contributions supressed by powers 
of the ratio of these two mass scales.

The aim of this paper is to apply the same idea to the case of an
interaction which does not require us to consider a nonrenormalizable
Lagrangian as a starting point. In this case, in contrast to the
gravitational interaction, it is not necessary to neglect any
contribution in order to identify an effective field theory 
with a finite number of independent parameters. One can 
interpret the present work as a step beyond the paradigm of
quantum field theory as a low-energy effective theory. Going
beyond the renormalizable theory (dominant term in the low-energy
limit) does not necessarily imply that we must consider an effective field 
theory  with an infinite number of free parameters. It is possible
to consider intermediate steps. The reduction of couplings, 
which could be a consequence of a symmetry of the underlying
fundamental theory which is hidden in the field theoretical limit,
corresponds to a situation where the field 
theoretical approach goes as far as possible in the sense that 
the low-energy limit of the theory is only sensitive to the 
details of the underlying theory through the value of a finite 
number of parameters. Instead of making reference to a symmetry
of a more fundamental theory, a reduction of couplings in an 
effective theory could be a consequence of the renormalization 
group flow in the infrared limit with a finite-dimensional
infrared fixed manifold. An interpretation along these lines
of the reduction of parameters in a renormalizable theory 
has been considered previously~\cite{Schrempp} in attempts
to determine the values of the Yukawa couplings in the 
standard model and in the minimal supersymetric standard model. 

All the ingredients in the discussion of the application of the 
method of reduction of couplings in an effective field theory
are present in the simplest case, the theory of a real scalar
field, which is the subject of this paper. The main result
of this work is the identification of a minimal extension of the 
renormalizable theory of a scalar field. This result is based on
a perturbative expansion of the renormalization group equations
for the renormalized parameters of the effective theory and 
therefore one can assume that it is a weakly interacting theory
over all the  range of validity of the energy expansion.

The triviality of the renormalizable scalar field theory, i.e.,
the impossibility to describe the interaction of scalar particles 
over an unlimited range of energies with a ${\phi}^{4}$ theory, 
is automatically incorporated when one considers the
theory as an effective field theory. As a consequence of the 
reduction of couplings it is possible to express all the 
corrections, which are proportional to inverse powers of the 
mass scale of the effective theory, in terms of a single 
additional parameter (together with 
the mass parameter and the quartic self-coupling of the
renormalizable ${\phi}^{4}$ theory). The extension of this result 
to the standard model of elementary particle physics considered
as an effective theory can have interesting physical applications 
if the characteristic mass scale of the theory is not much larger 
than the presently available energies.

In Sec.\ II  we consider the renormalization of the massless 
scalar field theory considered as an effective field theory. A
simple structure for the renormalization group equations is 
obtained in a mass-independent renormalization scheme due to
the absence of a dimensionfull ultraviolet cutoff. The 
renormalization group equation for the parameter corresponding 
to a term in the effective Lagrangian of a given dimension does
not depend on the parameters corresponding to terms of higher
dimension. It is this simple triangular structure which allows us
to find a solution to the renormalization group equations, 
where all the parameters corresponding to terms of dimension
greater than 4 in the Lagrangian can be expressed in terms 
of a single independent parameter $\lambda_{2}$, independently 
of the value of the renormalization scale. These relations 
between couplings can be uniquely determined order by order as 
an expansion in powers of the parameter $\lambda_{0}$ 
corresponding to the ${\phi}^{4}$ interaction.

In Sec.\ III the extension of the reduction of couplings to the
massive case is considered. Together with the expansion in powers
of the quartic scalar coupling $\lambda_{0}$ one has now an 
expansion in powers of the mass parameter $\lambda_{-2}$ in the 
renormalization group equations which gives corrections to
the triangular structure of the massless case. The reduction of 
couplings identified in Sec.\ II can be extended to this case
if one includes an expansion in powers of the product 
$\lambda_{-2}~\lambda_{2}$ in the relations between couplings and
one considers the mass parameter $\lambda_{-2}$ as an additional
independent parameter.

In Sec.\ IV the interpretation, limitations and some implications
of the effective scalar field theory after reduction of couplings
are discussed in detail. The physical content of the reduction
of couplings is disentangled from the presence of redundant terms
in the effective Lagrangian. A one-to-one correspondence 
between the three independent  parameters of the minimal extension 
of the renormalized scalar field theory  and three mass scales is
established. A hierarchy of mass scales in connection with the 
consistency of the perturbative reduction of couplings and also 
with the possible ambiguities induced by the high-order
behavior of the perturbative expansion (renormalons) is discussed
at the level of the effective theory. The standard study of the
effective potencial based on the renormalization group is generalized
to the case of an effective scalar field theory. The possibility
of spontaneous symmetry breaking and the modifications required in
the discussion of the renormalization and reduction of couplings
in the effective field theory are also considered. We end in 
Sec.\ V with a summary and prospects.

\section{Renormalization group equations and reduction of couplings: 
Massless case}

The starting point of our discussion is the more general expression 
for the effective Lagrangian of the theory of a real scalar field 
invariant under the discrete transformation $\phi\to~-\phi$.
It is convenient to introduce a fixed mass scale $M$ as a reference
unit for all the couplings of the effective theory. The effective 
Lagrangian can be written as an expansion in inverse powers of $M$: 

\begin{equation} {\cal L}_{eff} = \, 
{1\over 2}  \partial_{\mu} \phi
\partial^{\mu} \phi \, - \, {\lambda_{0}\over {4!}} \, \phi^{4} \, 
+ \, {{\vec \lambda}_{2} \over M^{2}} \, {\vec {\cal L}}^{(2)} \, 
+ \, {{\vec \lambda}_{4} \over M^{4}} \, {\vec {\cal L}}^{(4)} \, 
+ \, ...  \,\,. \label{Leff}  \end{equation}

\noindent A mass ($\phi^{2}$ term) has not been included (in next 
section we will see how the structure of the renormalization group
equations is affected in the presence of such a term). The coefficients 
$\lambda_{0}$, ${\vec \lambda}_{2}$, ${\vec \lambda}_{4}$,..., are 
dimensionless parameters and the power dependence on the mass scale 
$M$ is fixed by dimensional arguments. The effective field theory 
expansion has been written in a compact notation where 
${\vec {\cal L}}^{(2n)}$ is a vector whose components are the 
different terms of dimension $4 + 2n$ built out of the scalar field 
and its derivatives.

For the first terms in the effective field theory expansion one 
has

\begin{equation}
{\vec {\cal L}}^{(2)} \, = \, \left( {1 \over 6!} {\phi}^{6} \, , \,
{1 \over 4} {\phi}^{2} \partial_{\mu} \phi \partial^{\mu} \phi \, , \,
{1 \over 2} ({\Box \phi})^{2} \right) \,,\label{L2}
\end{equation}

\begin{equation}
{\vec {\cal L}}^{(4)} \, = \, \left({1 \over 8!} {\phi}^{8} \, , \,
{1 \over 2 (4!)} {\phi}^{4} \partial_{\mu} \phi \partial^{\mu} \phi 
\, , \, {1 \over 8} {(\partial_{\mu} \phi \partial^{\mu} \phi)}^{2}
\, , \,
{1 \over 2} {\phi} \partial_{\mu} \phi \partial^{\mu} \phi \Box \phi 
\, , \, 
{1 \over 4} {\phi}^{2} {(\Box \phi)}^{2} \, , \,
{1 \over 2} \Box \phi {\Box}^{2} \phi
\right) \,. \label{L4}
\end{equation}

The general parametrization of the effective Lagrangian is redundant 
for two different reasons. First, a change in the scale $M$ is
equivalent to an appropriate rescaling of every dimensionless
parameter. A choice of the scale $M$ such that all the dimensionless
parameters ${\lambda}_{2n}^{(i_n)}$ are simultaneously of order 1
allows us to identify $M$ with the scale characteristic of the energy
expansion of the effective theory. Second, by using a nonlinear 
redefinition of fields it is possible to eliminate some of the terms 
in the effective Lagrangian~\cite{Georgi,Arzt}. Nevertheless it
is simpler to use this redundant parametrization in order to
identify the general structure of the renormalization group
equations.

The standard derivation, in perturbatively renormalizable
theories, of the renormalization group equations in a 
mass-independent renormalization scheme\footnote[4]{See, for 
instance,~\cite{Gross}.} can be translated 
to an effective field theory. An infinite number of counterterms
must be admitted in order to absorb the infinities from loop
graphs~\cite{Zimmermann}. One can prove that limitations on the 
terms in the bare action arising from symmetries are compatible
with renormalizability~\cite{Gomis}. The effective theory  has 
an infinite number of bare
parameters in one-to-one correspondence with the dimensionless
parameters of the effective action. Using dimensional 
regularization one has expressions for the bare parameters
in terms of the renormalized parameters, with poles when
$\epsilon\rightarrow 0$ (dimension $D=4-\epsilon$).
>From the independence of the bare parameters on the 
renormalization scale $\mu$, one concludes that any change of
$\mu$ must be equivalent to a change in the renormalized
parameters. The renormalization group equations

\begin{equation}
\mu {d \lambda_{2n}^{(i_n)} \over d \mu} \,=\,
\beta_{{\lambda}_{2n}^{(i_n)}} ({\vec \lambda} \,)
\label{RGE} 
\end{equation}
 
\noindent express this fact. A straightforward generalization 
of the standard discussion of renormalizable theories~\cite{Gross}
leads to a perturbative determination of the renormalization
group $\beta$ functions from the residues of the simple poles at 
$\epsilon =0$ in the relations between bare and renormalized 
dimensionless parameters.

Dimensional arguments together with the presence of a single
mass scale $M$ (the dependence on the renormalization scale
$\mu$ is logarithmic) lead to the identification of a simple structure
for the renormalization group equations. The $\beta$ functions
satisfy the homogeneity conditions

\begin{equation}
\beta_{{\lambda}_{2n}^{(i_n)}} ({\vec \lambda}' \,) \, = \,
t^{2n} \, \beta_{{\lambda}_{2n}^{(i_n)}} ({\vec \lambda} \,)
\,, \label{hc2n} \end{equation}

\noindent where

\begin{equation}
{\vec \lambda}'_{2n} \, = \, t^{2n} \, {\vec \lambda}_{2n} 
\,. \label{ht} \end{equation}

These conditions put strong restrictions on the dependence 
of the $\beta$ functions on all the dimensionless parameters
with one exception, the scalar self coupling $\lambda_{0}$.
Each renormalization group $\beta$ function will be a polynomial
of a given degree in the parameters ${\vec \lambda}_{2n}$, 
$n\neq 0$, with coefficients which are a series expansion in 
$\lambda_{0}$ determined order by order in perturbation theory.

The explicit form of the renormalization group
equations for the first terms in the expansion of the
effective Lagrangian is given by

\begin{equation}
\mu {d \lambda_{0} \over d \mu} \,=\,
\lambda_{0}^{2} B_{0}  \,, \label{RGEl0} 
\end{equation}

\begin{equation}
\mu {d \lambda_{2}^{(i_{1})} \over d \mu} \, = \,
\lambda_{0} B_{2}^{(i_{1},j_{1})} \lambda_{2}^{(j_{1})}
\,, \label{RGEl2} 
\end{equation}

\begin{equation}
\mu {d \lambda_{4}^{(i_{2})} \over d \mu} \, = \,
\lambda_{0} B_{4}^{(i_{2},j_{2})} \lambda_{4}^{(j_{2})} \, + \,
B_{4}^{(i_{2},j_{1},k_{1})} \lambda_{2}^{(j_{1})} 
\lambda_{2}^{(k_{1})} \,, \label{RGEl4} 
\end{equation}

\noindent where the coefficients $B_{0}$, $B_{2}$ and $B_{4}$ are 
power expansions in the self-coupling $\lambda_{0}$. Indices
$i_{1}$, $j_{1}$, $k_{1}$ take three different values
corresponding to the three terms (\ref{L2}) in the effective 
Lagrangian of dimension 6 and $i_{2}$, $j_{2}$ distinguish
the six-dimensionless coefficients of terms of
dimension 8 (\ref{L4}). A sum over repeated indices 
$j_{1}$, $j_{2}$, $k_{1}$ is understoood in Eqs.\ (\ref{RGEl2}) 
and (\ref{RGEl4}). 

The $\mu$ dependence of ${\vec \lambda}_{2n}$ is fixed by a 
finite number of parameters ${\vec \lambda}_{2k}$ with
$k\leq n$. This triangular structure of the 
renormalization group equations allows a systematic search, 
order by order in the effective theory expansion, of relations
between the renormalized parameters independent of the
renormalization scale $\mu$ and compatible with the 
renormalization group equations, i.e., a reduction of couplings. 
In this way one can consider 
the possibility to have a finite number of independent
renormalized parameters despite the appearance of an infinite
number of interaction terms in the effective Lagrangian.

The first step in the reduction of couplings is to introduce 
a dimensionless parameter $\lambda_{2}$ with a renormalization 
scale dependence given by

\begin{equation}
\mu {d \lambda_{2} \over d \mu} \, = \,
\lambda_{0} B_{2} \lambda_{2}
\label{RRGEl2} \,,
\end{equation}

\noindent where the coefficient $B_{2}$ is an expansion in 
powers of $\lambda_{0}$ to be fixed in order to be able to 
write all the parameters $\lambda_{2}^{(i_{1})}$
in terms of $\lambda_{2}$ in a way compatible with the 
renormalization group equations (\ref{RGEl2}). The reduction
of couplings at this level corresponds to looking for a 
relation

\begin{equation}
\lambda_{2}^{(i_{1})} \, = \, \ell_{2}^{(i_{1})} \lambda_{2}
\label{REl2} \,.
\end{equation}

\noindent Consistency with the renormalization group equations
leads to 

\begin{equation}
B_{0} \lambda_{0} {d \ell_{2}^{(i_{1})} \over d \lambda_{0}} \, + \,
B_{2} \ell_{2}^{(i_{1})} \, = \, B_{2}^{(i_{1},j_{1})} 
\ell_{2}^{(j_{1})} \,, \label{CRl2} 
\end{equation}

\noindent which is a system of equations for the coefficients of
the reduction of couplings $\ell_{2}^{(i_{1})}$ and the coefficient
$B_{2}$ in the $\beta$ function of the independent
parameter $\lambda_{2}$. A loop expansion corresponds to a 
determination in perturbation theory of the $\beta$ functions
in Eq.\ (\ref{RGE}) and then to a determination of $B_{0}$ and
$B_{2}^{(i_{1},j_{1})}$ order by order as an expansion in
powers of $\lambda_{0}$:

\begin{eqnarray}
B_{0} \, & = & \, \sum_{k=0}^{\infty}
B_{0}^{(k)} \lambda_{0}^k
\ , \\
B_{2}^{(i_{1},j_{1})} \, & = & \, \sum_{k=0}^{\infty}
B_{2}^{(i_{1},j_{1};k)} \lambda_{0}^k
\ . \label{PEB}
\end{eqnarray}

\noindent The perturbative expansion of the 
renormalization group equations leads to a solution of the
consistency equations (\ref{CRl2}) with $\ell_{2}^{(i_{1})}$ and
$B_{2}$ given as an expansion in powers of $\lambda_{0}$:

\begin{eqnarray}
\ell_{2}^{(i_{1})} \, & = & \, \sum_{k=0}^{\infty}
\ell_{2}^{(i_{1} ; k)} \lambda_{0}^k
\,, \label{PEl2}\\
B_{2} \, & = & \, \sum_{k=0}^{\infty}
B_{2}^{(k)} \lambda_{0}^k
\,, \label{PEB2}
\end{eqnarray}

\noindent i.e., a perturbative determination of the reduction of
couplings. In lowest order, the consistency equation (\ref{CRl2})
reduces to

\begin{equation}
B_{2}^{(0)} \, \ell_{2}^{(i_{1};0)} \, = \, 
B_{2}^{(i_{1},j_{1};0)} \, \ell_{2}^{(j_{1};0)}
\,. \label{CRl20}
\end{equation}

For each eigenvector with a real eigenvalue 
of the matrix of lowest-order coefficients in the renormalization 
group equation of $\lambda_{2}^{(i_{1})}$ there is a consistent
reduction of these parameters. To lowest order in an expansion
in powers of $\lambda_{0}$, the coefficients of the
independent parameter $\lambda_{2}$ in the reduction equation
(\ref{CRl2}) are the components of the eigenvector, and the 
coefficient of the $\beta$ function of $\lambda_{2}$ is the 
corresponding eigenvalue. An extension of the reduction of couplings 
order-by-order in $\lambda_{0}$ leads to an order-by-order 
approximation of the consistency equation (\ref{CRl2}) which
reduces to a linear  system of equations for
the coefficients $\ell_{2}^{(i_{1} ; k)}$ of the reduction at
each order. 

A diagrammatic analysis allows us to identify easily the order in the
$\lambda_{0}$ expansion of the first term for each renormalization
group coefficient $B_{2}^{(i_{1},j_{1})}$. Just with this 
information it is possible to identify three different reductions
of couplings.

(1) In the first solution the three terms of dimension 6
in the effective Lagrangian have coefficients which begin at the same 
order in the $\lambda_{0}$
expansion. The coefficients in the reduction of couplings are 
determined in lowest order, up to an overall normalization factor
which can be reabsorbed into a redefinition of the independent
parameter $\lambda_{2}$. They are given by
\begin{eqnarray}
\ell_{2}^{(1;0)} \, &=& \, 1 \,, \label{R11}\\
\ell_{2}^{(2;0)} \, &=& \, B_{2}^{(2,1;0)} \, /
\left( B_{2}^{(1,1;0)} \, - \, B_{2}^{(2,2;0)} \right) 
\,, \label{R12}\\
\ell_{2}^{(3;0)} \, &=& \, B_{2}^{(3,2;0)} \, B_{2}^{(2,1;0)} \, / 
\left[B_{2}^{(1,1;0)} \left( B_{2}^{(1,1;0)} \, - \, B_{2}^{(2,2;0)} \right)
\right] \,. \label{R13}
\end{eqnarray}

\noindent The extension of the reduction of couplings to all
orders is uniquely determined once the arbitrariness in the
choice of the independent parameter is used to have
$B_{2}=B_{2}^{(1,1;0)}$ for the renormalization group 
coefficient of the independent parameter $\lambda_{2}$.

(2) A second solution has only terms with derivatives of 
the field in ${\cal L}_{2}$ in lowest order:

\begin{eqnarray}
\ell_{2}^{(1;0)} \, &=& \, 0 \,, \\
\ell_{2}^{(2;0)} \, &=& \, 1 \,, \label{R2}\\
\ell_{2}^{(3;0)} \, &=& \, B_{2}^{(3,2;0)} \, / 
B_{2}^{(2,2;0)}\,, 
\end{eqnarray}

\noindent and $B_{2}=B_{2}^{(2,2;0)}$.

(3) The last solution has only the term with four
derivatives to start with,

\begin{equation}
\ell_{2}^{(1;0)} \, = \, \ell_{2}^{(2;0)} \, = \, 0
\,, \, \, \, \, \, \,  \ell_{2}^{(3;0)} \, = \, 1 \,,\label{R3}
\end{equation} 

\noindent and the $\beta$ function of the independent
parameter is proportional to $\lambda_{0}^{2}$ in this case:

\begin{equation}
B_{2} \, = \, \left( B_{2}^{(3,3;1)} \, - \,
{B_{2}^{(3,2;0)} B_{2}^{(2,3;1)} \over B_{2}^{(2,2;0)}}
\right) \, \lambda_{0} \,.
\label{B2R3}
\end{equation}

Once the reduction of couplings at the level of terms of 
dimension 6 in the effective Lagrangian has been 
implemented, the next step is to consider the renormalization
group equation for the coefficients of terms of dimension 8.
Using the reduction of couplings (\ref{REl2}) one has

\begin{equation}
\mu {d \lambda_{4}^{(i_{2})} \over d \mu} \, = \,
\lambda_{0} B_{4}^{(i_{2},j_{2})} \lambda_{4}^{(j_{2})} \, + \,
L_{4}^{(i_{2})} \lambda_{2}^{2} 
\,, \label{RRGEl4} 
\end{equation}

\noindent where 

\begin{equation}
L_{4}^{(i_{2})} \, = \, B_{4}^{(i_{2},j_{1},k_{1})} \,
\ell_{2}^{(j_{1})} \, \ell_{2}^{(k_{1})} \,. \label{RRGE2l4}
\end{equation}

\noindent Now one has to look for the possibility to
express the parameters ${\vec \lambda}_{4}$ as a
function of $\lambda_{0}$ and $\lambda_{2}$ in such a
way that one reproduces their renormalization scale
dependence, given in Eq.\ (\ref{RRGEl4}), as a consequence
of the renormalization group equations
(\ref{RGEl0}), (\ref{RRGEl2}) 
of $\lambda_{0}$ and $\lambda_{2}$. A relation

\begin{equation}
\lambda_{4}^{(i_{2})} \, = \, 
{\ell_{4}^{(i_{2})} \over \lambda_{0}} \, \lambda_{2}^2 \,,
\label{REl4}
\end{equation}

\noindent where the coefficient $\ell_{4}^{(i_{2})}$ is a function
of $\lambda_{0}$, will be consistent with the renormalization
group equations if

\begin{equation}
B_{0} \lambda_{0} {d \ell_{4}^{(i_{2})} \over d \lambda_{0}} \, + \,
\left( 2 B_{2} - B_{0} \right) \ell_{4}^{(i_{2})} \, = \, 
B_{4}^{(i_{2},j_{2})} \ell_{4}^{(j_{2})} \, + \, L_{4}^{(i_{2})}
\,. \label{CRl4} 
\end{equation}

\noindent This is a system of equations for the coefficients in the 
reduction $\ell_{4}^{(i_{2})}$ and all other factors are expansions 
in powers of $\lambda_{0}$ which are determined order by order
either directly from the perturbative approximation to the 
renormalization group equations or from the perturbative determination
of the reduction of couplings at the previous level. A solution of 
the consistency equations (\ref{CRl4})  with $\ell_{4}^{(i_{2})}$ 
given as an expansion in powers of $\lambda_{0}$, 

\begin{equation}
\ell_{4}^{(i_{2})} \, = \, \sum_{k=0}^{\infty}
\ell_{4}^{(i_{2} ; k)} \lambda_{0}^k
\,, \label{PEl4}
\end{equation}

\noindent is obtained by solving a linear system of equations at each 
order in the expansion in powers of $\lambda_{0}$
of the consistency equations.

The steps followed in the determination of the reduction of the
parameters corresponding to terms of order $1/M^4$ can
be repeated order by order in the expansion in $1/M$ to get
the reduction of the effective scalar field theory. 
It is given by the relations 

\begin{equation}
\lambda_{2n}^{(i_{n})} \, = \, 
{\ell_{2n}^{(i_{n})} \over \lambda_{0}^{n-1}} \, \lambda_{2}^n \,,
\label{REl2n}
\end{equation}

\noindent where the coefficients $\ell_{2n}^{(i_{n})}$ are expansions
in powers of $\lambda_{0}$,

\begin{equation}
\ell_{2n}^{(i_{n})} \, = \, \sum_{k=0}^{\infty}
\ell_{2n}^{(i_{n} ; k)} \lambda_{0}^k
\,, \label{PEl2n}
\end{equation}

\noindent determined by the renormalization group equations.

The final result is an effective Lagrangian with an infinite
number of terms of higher dimension added to the massless
renormalizable scalar field Lagrangian but with only one
additional independent renormalized parameter $\lambda_{2}$,
with a renormalization scale dependence determined by a
one loop calculation. In fact we have found three 
different minimal extensions of the renormalizable theory of 
this kind. 

The reduction of couplings should not be confused with the 
identification of redundant terms in the effective Lagrangian.
By using a nonlinear redefinition of fields it is possible
to eliminate all the terms of dimension greater than 4
involving $\Box\phi$~\cite{Georgi,Arzt}. For example, by
making the shift of variables $\phi\to\phi^{'}$ with

\begin{equation}
\phi \, = \, \phi^{'} \, + \, \left( {\lambda_{2}^{(2)}\over M^2} -
{1 \over 2} \lambda_{0} {\lambda_{2}^{(3)}\over M^2} \right) 
{{\phi^{'}}^3 \over 3!}
- {1 \over 2} {\lambda_{2}^{(3)}\over M^2} \, \Box \phi^{'} \,,
\label{phi'}
\end{equation}

\noindent one has a Lagrangian with only one term of dimension 6,
${\hat \lambda}_{2}/M^2\,(1/6!)\,{\phi^{'}}^6$, where

\begin{equation}
{\hat \lambda}_{2} \, = \, \lambda_{2}^{(1)} \, - \,
20 \, \lambda_{0} \, \lambda_{2}^{(2)} \, + \, 10 \, 
\lambda_{0}^2 \, \lambda_{2}^{(3)} \,.
\label{l2hat}
\end{equation}

\noindent Then at this level the simplification of the effective 
Lagrangian due to the presence of redundant terms has a similar 
effect as the reduction of couplings but this is not the case if
one considers higher-dimensional terms. If one includes higher
dimensional terms in the change of variables it is possible to
extend the simplification of the effective Lagrangian to terms 
of dimension higher than 6. At the $1/M^4$ level it
is possible to eliminate three out of the six terms in Eq.\ (\ref{L4})
but one still has three new independent parameters to be
compared with the absence of any additional free parameters after
reduction of couplings.

\section{Renormalization group equations and reduction of couplings: 
Massive case}

If one considers a massive spinless particle then one has to include 
a term $\lambda_{-2} M^{2}\phi^{2}$ in the Lagrangian density.
The dimensionless parameter $\lambda_{-2}$ has to be taken into
account in the discussion based on dimensional arguments leading 
to the general structure of the renormalization group equations.
The homogeneity conditions of the $\beta$ functions 
include the rescaling of the additional parameter
$\lambda^{'}_{-2}=t^{-2}\,\lambda_{-2}$ and the simple
triangular structure is lost due to the contributions proportional
to positive powers of $\lambda_{-2}$ which will be accompanied by 
parameters corresponding to terms of higher dimensionality. If 
one wants the reduction of couplings to be applicable also in this 
case then one has to assume that the dimensionless parameter
$\lambda_{-2}$ is sufficiently small to treat its effects as a
small perturbation.  The reduction of couplings identified in
the previous section for the massless case can be taken as the 
zero-order term of an expansion of the reduction equations
in powers of the parameter $\lambda_{-2}$.

The renormalization group equation for the dimensionless
parameter associated to the mass term is given by

\begin{eqnarray}
 \mu {d \lambda_{-2}\over d \mu} \, &= &\, B_{-2,0} \lambda_{-2} 
\lambda_0 + \left[  B_{-2,1}^{(i_1)} \lambda_2^{(i_1)} \right] 
{\lambda_{-2}}^2 + \nonumber \\
& & \left[  B_{-2,2}^{(i_2)} \lambda_4^{(i_2)} +
B_{-2,2}^{(i_1,j_1)} \lambda_2^{(i_1)} \lambda_2^{(j_1)} \right]
{\lambda_{-2}}^3 + \cdots \, \,,\label{RGEl-2}
\end{eqnarray}

\noindent where the coeficients $B_{-2,k}$ are power expansions
in $\lambda_{0}$ determined from a perturbative calculation 
of counterterms. As a consequence of the homogeneity conditions,
$\beta_{\lambda_{-2}}$ is proportional to $\lambda_{-2}$ and then 
a vanishing mass parameter considered in Sec.\ II is consistent
with the renormalization group equations. For the self-coupling 
$\lambda_{0}$ one has

\begin{eqnarray}
 \mu {d \lambda_0\over d \mu} \, &=&\,  B_{0,0}\lambda_0^2 + 
   \left[  B_{0,1}^{(i_1)} \lambda_2^{(i_1)} \right] {\lambda_{-2}} + 
   \nonumber \\
 & & \left[  B_{0,2}^{(i_2)} \lambda_4^{(i_2)} +
    B_{0,2}^{(i_1,j_1)} \lambda_2^{(i_1)} \lambda_2^{(j_1)} \right]
   {\lambda_{-2}}^2 + \cdots \, \,.\label{RGEl02}
\end{eqnarray}

\noindent The first term is just the massless $\beta$ function since
$B_{0,0}$ is just the coefficient $B_{0}$ of the massless renormalization
group equation. One has additional terms proportional to positive powers
of $\lambda_{-2}$ with coefficients $B_{0,k}$ determined perturbatively.
The renormalization scale dependence of the parameters corresponding
to terms of dimension 6, which in the massless case was given by
Eq.\ (\ref{RGEl2}), will now take the form

\begin{eqnarray}
 \mu {d \lambda_{2}^{(i_1)}\over d \mu} \, = \, B_{2,0}^{(i_1,j_1)}
   \lambda_{2}^{(j_1)}\lambda_0 + 
 \left[ B_{2,1}^{(i_1,i_2)} \lambda_4^{(i_2)} +
   B_{2,1}^{(i_1,j_1,k_1)} \lambda_2^{(j_1)} \lambda_2^{(k_1)} \right]
   {\lambda_{-2}} + \nonumber \\
 \left[ B_{2,2}^{(i_1,i_3)} \lambda_6^{(i_3)} +
   B_{2,2}^{(i_1,j_1,i_2)} \lambda_2^{(j_1)} \lambda_4^{(i_2)} + 
B_{2,2}^{(i_1,j_1,k_1,\ell_1)} \lambda_2^{(j_1)}
   \lambda_2^{(k_1)} \lambda_2^{(\ell_1)} \right] {\lambda_{-2}}^2 + 
\cdots \, \,.\label{RGEl22} 
\end{eqnarray}

Equations (\ref{RGEl-2})--(\ref{RGEl22}), toghether with its
obvious generalization for the remaining parameters in the 
effective Lagrangian, are the starting point for an extension
to the massive case of the reduction of couplings discussed in the
previous section. The presence of a new independent parameter
$\lambda_{-2}$ and the general structure of the renormalization 
group equations leads to the consideration in the massive case of a
relation

\begin{equation}
 \lambda_2^{(i_1)} = \left[ \ell_{2,0}^{(i_1)}  + 
 \ell_{2,1}^{(i_1)} \,  {{\lambda_2 \lambda_{-2}} \over \lambda_{0}^2} 
+ \ell_{2,2}^{(i_1)} \, \left({{\lambda_2 \lambda_{-2}} \over 
\lambda_{0}^2} \right)^2 + \cdots 
\right] \lambda_2 \,, \label{REl22}
\end{equation}

\noindent fixing the effective Lagrangian at order $1/M^2$ 
in terms of the parameters $\lambda_{-2}$, $\lambda_{0}$, and 
$\lambda_{2}$. The coeficients $\ell_{2,k}$ are determined by the 
consistency of Eq.\ (\ref{REl22}) with the renormalization group equations. 
At each order in the expansion in powers of $\lambda_{-2}$ one has a 
system of equations for the coefficients of the reduction of couplings. 
In lowest order one has the massless consistency equation (\ref{CRl2})
for $\ell_{2,0}$. At order $\lambda_{-2}$ the consistency of the 
reduction of $\lambda_{2}^{(i_{1})}$ with the renormalization group
equations leads to 

\begin{eqnarray}
{d \ell_{2,0}^{(i_1)} \over d \lambda_0}
 B_{0,1}^{(j_1)} \ell_{2,0}^{(j_1)} + \left( 2 B_{2,0}+B_{-2,0}\right)
\ell_{2,1}^{(i_1)} + {d \ell_{2,1}^{(i_1)} \over d \lambda_0} B_{0,0} 
= \nonumber \\
 B_{2,0}^{(i_1,j_1)} \ell_{2,1}^{(j_1)} + 
  B_{2,1}^{(i_1,i_2)} \ell_{4,0}^{(i_2)} +
  B_{2,1}^{(i_1,j_1,k_1)} \ell_{2,0}^{(j_1)} \ell_{2,0}^{(k_1)}
\,, \label{CRl22}
\end{eqnarray}

\noindent which determines the coefficients $\ell_{2,1}^{(i_1)}$ as an 
expansion in powers of $\lambda_{0}$ once the coefficient 
$\ell_{4,0}^{(i_2)}$ in the reduction of the parameter 
$\lambda_{4}^{(i_2)}$ has been determined from the consistency with the 
renormalization group equations in lowest order [Eq.\ (\ref{CRl4})]. 
This argument can be repeated step by step obtaining an effective 
Lagrangian with three independent parameters $\lambda_{-2}$, $\lambda_{0}$, 
and $\lambda_{2}$. The dimensionless coefficient of a generic term will be 
given by

\begin{equation}
 \lambda_{2n}^{(i_n)} = \left[ \ell_{2n,0}^{(i_n)}  + 
 \ell_{2n,1}^{(i_n)}  \,  {{\lambda_2 \lambda_{-2}} \over \lambda_{0}^2} +
 \ell_{2n,2}^{(i_n)} \, \left({{\lambda_2 \lambda_{-2}} \over 
\lambda_{0}^2} \right)^2 + \cdots
\right]  {{\lambda_2}^n \over \lambda_0^{n-1}} \ , \label{REl2n2}
\end{equation}

\noindent where $\ell_{2n,k}^{(i_n)}$ are power expansions in 
$\lambda_{0}$ determined by the consistency with the renormalization 
group equations. The determination of the reduction coefficients goes
from lower to higher values of $k$ (i.e., order by order in
the expansion in powers of $\lambda_{-2}$ of the renormalization
group equation), for a given value of $k$ it goes from lower
to higher values of $n$ (i.e., order by order in the effective
Lagrangian expansion), and for a given value of $n$ and $k$ it
goes order by order in perturbation theory ($\lambda_{0}$ expansion).
Once a solution for the first coefficients ($n\,=\,1$, $k\,=\,0$) in
lowest order has been obtained [Eqs.\ (\ref{R11})--(\ref{B2R3})] the 
reduction of the effective Lagrangian is determined by solving
linear  systems of equations for the remaining
coefficients of the reduction.

The effective theory is defined by the relations giving each
coefficient in the effective Lagrangian in terms of the three 
independent parameters and by the renormalization group equations
which give the renormalization scale dependence of the
independent parameters. The arbitrariness in the choice of the
independent parameter $\lambda_{2}$ has been used in order to have
a scale dependence given by Eq.\ (\ref{RRGEl2}), where $B_{2}$ is 
either a constant or a constant times $\lambda_{0}$ depending
on the solution to the lowest-order consistency equations. The
renormalization group equations for $\lambda_{0}$ and $\lambda_{-2}$,
which are the parameters of the renormalizable $\phi^{4}$ theory,
are given by

\begin{eqnarray}
 \mu {d \lambda_{-2}\over d \mu} \,& = &\, \left[ L_{-2,0} \, + \, 
L_{-2,1} \, {{\lambda_2 \lambda_{-2}} \over \lambda_{0}^2} +
L_{-2,2} \, \left({{\lambda_2 \lambda_{-2}} \over 
\lambda_{0}^2} \right)^2 + \cdots \right] \, \lambda_{0} \,
\lambda_{-2} \, , \\
 \mu {d \lambda_{0}\over d \mu} \,& = &\, \left[ L_{0,0} \, + \, 
L_{0,1} \, {{\lambda_2 \lambda_{-2}} \over \lambda_{0}^2} +
L_{0,2} \, \left({{\lambda_2 \lambda_{-2}} \over 
\lambda_{0}^2} \right)^2 + \cdots \right] \, \lambda_{0}^2 \, ,
\label{RRGE}
\end{eqnarray}

\noindent where the coeficients $L_{-2,k}$, $L_{0,k}$ are obtained by 
combining the renormalization group equations (\ref{RGEl-2}), 
(\ref{RGEl02}) with the reduction relations (\ref{REl2n2}). One 
has for the first coefficients

\begin{eqnarray}
L_{-2,0} \,& = &\, B_{-2,0} \, , \nonumber \\
L_{-2,1} \,& = &\lambda_{0}\, B_{-2,1}^{(i_1)} \, \ell_{2,0}^{(i_1)} 
\, , \nonumber \\
L_{-2,2} \,& = &\lambda_{0}\, B_{-2,1}^{(i_1)} \, \ell_{2,1}^{(i_1)} \, + \,
\lambda_{0}^2 \, B_{-2,2}^{(i_2)} \, \ell_{4,0}^{(i_2)} \, + \,
\lambda_{0}^3 \, B_{-2,2}^{(i_1,j_1)} \, \ell_{2,0}^{(i_1)} 
\ell_{2,0}^{(j_1)} \, , \nonumber \\
L_{0,0} \,& = &\, B_{0,0} \, , \nonumber \\
L_{0,1} \,& = &\, B_{0,1}^{(i_1)} \, \ell_{2,0}^{(i_1)} \, , \nonumber \\
L_{0,2} \,& = &\, B_{0,1}^{(i_1)} \, \ell_{2,1}^{(i_1)} \, + \,
\lambda_{0} \, B_{0,2}^{(i_2)} \, \ell_{4,0}^{(i_2)} \, + \,
\lambda_{0}^2 \, B_{0,2}^{(i_1,j_1)} \, \ell_{2,0}^{(i_1)} \,
\ell_{2,0}^{(j_1)} \, , \nonumber
\end{eqnarray}

\noindent and then one reproduces the renormalization group 
equations of the renormalizable $\phi^{4}$ theory plus 
corrections due to the extension which are determined
perturbatively.  

\section{Some aspects of the effective scalar field theory after 
reduction of couplings}

In order to discuss the properties of a scalar field theory considered
as a low-energy effective theory with three free parameters it is 
convenient to introduce a mass scale associated to each of the
independent parameters. For the self-coupling $\lambda_{0}$ one can 
consider the approximation to the renormalization group equation where
the corrections proportional to $\lambda_{-2}\lambda_{2}$ are neglected,
i.e., the renormalization group equation of the $\phi^{4}$ theory, and at
this level one can identify the scale $M_{0}$ at which the perturbative
approach breaks down (Landau pole). 

Associated to the parameter 
$\lambda_{2}$, which controls the departure from the renormalizable
$\phi^{4}$ theory, one can consider a new scale $M_{2}$. A comparison of the
lowest-order contribution to the $2\to 4$ scattering cross section in the
renormalizable $\phi^4$ theory with the first contribution from the 
higher-dimensional terms in the effective Lagrangian can be used to define
the scale $M_{2}$ as the energy where both contributions become
comparable. That leads to the identification of $M_{2}$ as the scale $M$ in the 
effective Lagrangian (\ref{Leff}) 
such that 

$${\hat \lambda}_{2}(\mu=M_2)\,=\, 
\lambda_{0}^2(\mu=M_2).$$ 

\noindent Note that the scale $M_{2}$ is defined through the parameter
${\hat \lambda}_{2}$ which is the coefficient of the $\phi^6$ term in
the effective Lagrangian after a shift of variables has been made to
eliminate the remaining terms of dimension 6. Then all the arbitrariness 
in the parametrization of the effective Lagrangian cancel in the 
determination of $M_{2}$, as should be since it can be taken 
as a measure of the energy range where the effective theory expansion 
is a good aproximation. 

The third scale $M_{-2}$ gives a first aproximation to the mass of the 
spinless particle. It can be introduced by the condition 
$$\lambda_{-2}(\mu=M_{-2})M_{2}^{2}\,=\,M_{-2}^{2}$$ 
\noindent on the
coefficient  $\lambda_{-2}$ of the $\phi^2$ term in the effective
Lagrangian for $M=M_{2}$. In the determination of the scale 
$M_{-2}$, as a function of the parameters $\lambda_{-2}$ and $\lambda_{0}$
at the scale $\mu =M_{2}$, the terms proportional to $\lambda_{2}$
in the renormalization group equation of $\lambda_{-2}$ are neglected.

\subsection{Limitations of the minimal extension of the renormalizable
$\phi^4$ theory}

The renormalizability of the scalar field theory with three independent
parameters has been discussed order by order in a perturbative expansion 
in the self-coupling $\lambda_{0}$. Therefore, unless a generalization of this 
result at the nonperturbative level is found, one has to assume that 
$\lambda_{0}(\mu)$ is smaller than the value of the coupling at which
perturbation theory becomes unreliable for any scale in the range of 
validity of the effective theory. There are ambiguities
in the determination of this value, which in the case of the 
$\phi^4$ theory leads to the identification of
$\lambda_{0}\approx$ 3--4 as the value 
at which the theory becomes strongly interacting.\footnote[5]{For a recent 
discussion, see~\cite{Willenbrock}.}
The conclusion is that the scale $M_{2}$ 
limiting the range of validity of the effective theory expansion can not 
exceed the scale $M_{0}$ associated to the parameter $\lambda_{0}$. More
precisely one has the condition 
$${{\lambda_{0}(\mu = M_{2})} \over {16 \pi^2}} \leq \epsilon_{0}$$ 
\noindent where $\epsilon_{0}$ fixes the domain
of validity of the  perturbative expansion using several
criteria~\cite{Willenbrock}  (suppression of higher-order terms, decrease of
renormalization scale  dependence, absence of significant violations of
unitarity,...).
A second obvious limitation is that one can only consider low-energy
observables such that the ratio $E^2/M_{2}^2$ is small enough 
to justify the use of the effective Lagrangian expansion. 

The third limitation comes from the expansion in the reduction of
couplings due to the introduction of a mass term in the effective
Lagrangian. The validity of the step by step reduction of couplings 
requires that 
$${\lambda_{2}\lambda_{-2} \over \lambda_{0}^2}\leq\epsilon_{2}$$ 
\noindent over all the
energy range of validity of the  effective theory. In order to translate this
condition into a limitation on the mass scales of the effective theory we have
to use the  renormalization group equation for the independent parameters and
the  explicit form of the reduction of couplings. 

For definiteness we consider 
the first reduction, Eqs.\ (\ref{R11})--(\ref{R13}), identified in Sec.\ II. 
In this
case,  neglecting higher-order terms in the $\lambda_{0}$ expansion, one has 
${\hat \lambda}_{2}=\lambda_{2}$ and, as a consequence of the 
definition of the scale $M_{2}$, $\lambda_{2}(M_{2})=
\lambda_{0}^2(M_{2})$. The consistency of the step-by-step
reduction leads to the condition $\lambda_{-2}(M_{2})\leq
\epsilon_{2}$. A one loop calculation determines the lowest-order 
approximation to the renormalization group equations for the independent
parameters which, in the case of the reduction in Eqs.\ 
(\ref{R11})--(\ref{R13}), reads
\begin{eqnarray}
\mu {d \over d \mu} \left( {\lambda_{0} \over {16 \pi^2}} \right)
\,& = &\, 3 \, \left( {\lambda_{0} \over {16 \pi^2}} \right)^2
\,, \\
\mu {d \lambda_{-2} \over d \mu} \,& = &\, \lambda_{-2} \,
\left( {\lambda_{0} \over {16 \pi^2}} \right) \,,\label{RRGE02} \\
\mu {d \lambda_{2} \over d \mu} \,& = &\, 15 \,\, \lambda_{2} 
\left( {\lambda_{0} \over {16 \pi^2}} \right) \,.
\end{eqnarray}

\noindent Taking as a reference the parameters at the scale
$M_{2}$, one has a renormalization scale dependence for the 
self-coupling given by

\begin{equation}
\lambda_{0} ( \mu ) \, = \, {\lambda_{0} ( M_{2} ) \over
1 \, + \, 3/2 [\lambda_{0} ( M_{2} )/16 \pi^2] \,
\ln \left(M_{2}^2/\mu^2\right) } \,.\label{l0}
\end{equation}

\noindent Then, in this approximation, one has a simple 
expression for the ratio of scales $M_{0}^2 / M_{2}^2$
in terms of the parameter $\lambda_{0}(M_{2})$:

\begin{equation}
{M_{0}^2 \over M_{2}^2} \, = \, \exp \left[{2 \over 3} 
{1 \over \lambda_{0} ( M_2 ) / 16 \pi^2} \right]
\,.\label{M0M2}
\end{equation}

\noindent The solution of the renormalization group equations for
the parameters $\lambda_{2}$, $\lambda_{-2}$ is

\begin{eqnarray}
\lambda_{-2} ( \mu ) \,& = &\, \lambda_{-2} ( M_{2} )  
\left[ {\lambda_{0} ( \mu ) \over \lambda_{0} ( M_{2} )}
\right]^{1/3} \,, \\
\lambda_{2} ( \mu ) \,& = &\, \lambda_{2} ( M_{2} )  
\left[ {\lambda_{0} ( \mu ) \over \lambda_{0} ( M_{2} )}
\right]^5 \,,\label{prueba}
\end{eqnarray}

\noindent and then one has, for the combination of parameters which 
appears in the expansion of the reduction of couplings of the
massive case,

\begin{equation}
{\lambda_{2} ( \mu )  \lambda_{-2} ( \mu )  \over 
\lambda_{0}^2 ( \mu )} \, = \, 
{\lambda_{2} ( M_{2} )  \lambda_{-2} ( M_{2} )  \over 
\lambda_{0}^2 ( M_{2} )} \, 
\left[ {\lambda_{0} ( \mu ) \over \lambda_{0} ( M_{2} )}
\right]^{10/3} \,.\label{MER}
\end{equation} 

\noindent Since the coupling $\lambda_{0}$ decreases when one 
goes to lower scales, the convergence of the expansion of the reduction 
of couplings over all the energy range of validity of the effective 
theory is automatic once the couplings at the scale $M_{2}$ has been 
chosen appropriately $[\lambda_{-2}(M_2)\leq\epsilon_2]$.

One can also use the explicit form of the solution of the renormalization 
group equations to translate the restriction 
$\lambda_{-2}(M_{2})\leq\epsilon_{2}$ into a restriction on the
ratio of mass scales $M_{-2}^2 / M_{2}^2$:

\begin{equation}
\left[ 1 \, + \, 
{3 \over 2} {\lambda_{0} ( M_{2} ) \over {16 \pi^2}} 
\ln \left({M_{2}^2 \over M_{-2}^2}\right)  \right]^{1/3} \,
{M_{-2}^2 \over M_{2}^2} \, \leq \, \epsilon_{2}
\,.\label{BM-2}
\end{equation}

There is a clear correspondence between the limitations on the scales
of the effective theory and the different bounds obtained in the 
$\phi^4$ theory. The limitations of perturbation theory in the 
$\phi^4$ theory, including the perturbative unitarity bound, are
automatically incorporated in the perturbative approach to the 
effective Lagrangian and the treatment of the higher-dimensional
terms as a small perturbation implies that the perturbative bounds
in the effective theory will be very close to the bounds of the
$\phi^4$ theory. 

With respect to the triviality bounds, these are
usually formulated as a restriction on the renormalized parameters
due to the neccessity of a finite cutoff $\Lambda$ in order to have
a nontrivial interacting system. A lattice formulation of the 
$\phi^4$ theory leads~\cite{Luscher} to the identification of an upper bound on 
the scalar mass in units of the cutoff and an upper bound on the 
renormalized coupling if one limits the size of the deviations 
from continuum theory. As long as the higher-dimensional terms in 
the effective Lagrangian are a small perturbation, the result that
the bound on the coupling is smaller than the perturbative bound and 
the conclusion that there is no strongly interacting theory, can be 
translated to the effective theory justifying its perturbative 
treatment. It is not clear whether the modification on the perturbative
bound for the coupling due to the higher-dimensional terms could be big 
enough within the domain of validity of the effective theory expansion
to make possible a strongly interacting theory.

There is a relation between scaling violations in lattice $\phi^4$ theory
and deviations of the effective theory from the $\phi^4$ theory
which can be obtained if one uses the local effective Lagrangian
description of scaling violations~\cite{Symanzik}. An Euclidean
lattice $\phi^4$ theory with a given lattice action has the same
perturbative expansion as an effective scalar theory with a
Lagrangian given by the local effective Lagrangian which describes
the scaling violations of the lattice theory.
The upper bound on the mass in units of the cutoff obtained in
the lattice field theory analysis can be translated 
to the effective scalar theory if one identifies the 
scale $M_{2}$, which is a measure of the domain of validity of the 
effective theory expansion, with the cutoff of the lattice 
$\phi^4$ theory. The bounds in the lattice theory on the deviations
from the continuum (scaling violations) are associated to the bounds 
in the effective theory on the deviations from the $\phi^4$ theory. 

Another possible source of limitations of the perturbative treatment of
the effective theory is the divergence of the perturbation series. A
comparison of the $n$th order term in the $\lambda_{0}$ expansion with
the first correction due to higher-dimensional terms leads to the ratio

\begin{equation}
{\left(\lambda_{0} / 16 \pi^2 \right)^n \over
\lambda_{2} \lambda_{-2} / \lambda_{0}^2} \, ( \mu ) \, = \,
{\left(\lambda_{0} / 16 \pi^2 \right)^n \over
\lambda_{2} \lambda_{-2} / \lambda_{0}^2} \, ( M_{2} ) \,
\left[ {\lambda_{0} ( \mu ) \over 16 \pi^2} \right]^{n - 10/3}
\label{n}
\end{equation}

\noindent which, for any given scale $\mu$, becomes smaller than 1
for $n$ sufficiently large. Larger values of the scale require us to go to
higher orders in order to have a perturbative correction smaller than
the contribution due to higher-dimensional terms. This means that it
makes no sense to worry about the large-order behavior of the 
perturbation series while neglecting the higher-dimensional terms in
the effective Lagrangian.

Assuming that the large-order behavior of perturbation 
theory is dominated by the renormalon singularity~\cite{Renormalon}
leads to an ambiguity in the sum of the perturbation series which
is of order $E^2 / M_{0}^2$. Although there is no physical 
significance to these ambiguities when treated 
consistently~\cite{Manohar2} still one can describe the effect of a 
truncation in the perturbative expansion by these 
ambiguities~\cite{Martinelli}. If one has an effective theory with
$M_{2}^2 / M_{0}^2 \ll 1$, i.e., if the self-coupling
at the scale $M_{2}$ is such that $\lambda_{0}(M_{2})
/ 16\pi^2 \ll 1$, then the corrections of order 
$E^2 / M_{2}^2$ due to higher-dimensional terms are much bigger
than the ambiguities due to the divergences of the perturbation 
series. On the contrary if one considers an effective 
theory where $M_{0}$ and $M_{2}$ are of the same 
order then the ambiguities due to the divergence in the perturbation
series are of the same order as the corrections to the 
$\phi^4$ theory.\footnote[6]{Similar conclusions are obtained from a 
different point of view in~\cite{Espriu}.} Once more there is a 
correspondence between the previous discussion of renormalons in the 
effective scalar field theory and the connection between scaling 
violations in lattice $\phi^4$ theory and the divergence of the 
perturbation series~\cite{Luscher}.
To end this section, let us remark that, although the first of the
three possible reductions of couplings of the scalar theory was
used in the discussion of the limitations in the effective theory, 
similar arguments can be used for the other cases of reduction of couplings.

\subsection{Physical content of the reduction of couplings}

The cross section for any process in the scalar theory can be written
using simple dimensional analysis in the form

\begin{equation}
\sigma \, = \, {1 \over E^2} {\hat \sigma} 
\left( x, \lambda_{0} ( \mu ) , {\lambda_{-2} ( \mu ) M_{2}^2 \over E^2} ,
{\lambda_{2} ( \mu ) E^2 \over M_{2}^2} , {\mu^2 \over E^2} \right) \,,
\label{cs}
\end{equation}

\noindent where $E$ is an overall energy scale of the process, $x$ denotes
angles and energy ratios and $\mu$ is the renormalization scale. The 
independence of the cross section $\sigma$ on the  renormalization scale
can be used to choose $\mu =E$.

In order to apply the effective Lagrangian expansion one has to consider 
the scattering of jets instead of particles in order to have cross 
sections which remain finite when the mass vanishes. Using the 
renormalized parameters corresponding to a mass-independent 
renormalization scheme, as has been done in the discussion of the
renormalization group equations of the effective theory leading to
the reduction of couplings, it is possible to expand any cross section
in powers of the independent dimensionless parameters. One has

\begin{equation}
{\hat \sigma} \, = \, \sum_{i,j=0}^{\infty} {\hat \sigma}^{(i,j)} 
\left( \lambda_{0} ( E ) , x \right) 
\left( {\lambda_{-2} ( E ) M_{2}^2 \over E^2} \right)^i 
\left( {\lambda_{2} ( E ) E^2 \over M_{2}^2} \right)^j \,,
\label{csexp}
\end{equation}

\noindent where the coefficients ${\hat \sigma}^{(i,j)}$ can be
determined order by order as an expansion in powers of $\lambda_{0}$
from the perturbative calculation of the cross section and the
reduction of couplings. Using the renormalization group equations
for the independent parameters and the definition of the mass 
scales of the effective theory one can rewrite the cross section
in the form

\begin{equation}
\sigma \, = \, {1 \over E^2} \sum_{i,j=0}^{\infty} {\hat \sigma}^{(i,j)} 
\left( \lambda_{0} ( E ) , x \right) 
\left[ {\lambda_{-2} ( E ) \over \lambda_{-2} ( M_{-2} )}
\right]^i
\left[ \lambda_{2} ( E ) \right]^j
\left( {M_{-2}^2 \over E^2} \right)^i
\left( {E^2 \over M_{2}^2} \right)^j \,,
\label{csexp2}
\end{equation}

\noindent i.e., as a double expansion in the ratio of the mass of 
the particle over the energy of the process $M_{-2}^2 / E^2$
and the ratio of the energy of the process over the scale $M_{2}$.

As a consequence of the reduction of couplings in the effective 
theory it is possible to get a systematic approximation to any
cross section in terms of a mass scale $M_{2}$, a self-coupling
$\lambda_{0}(M_{2})$, and $\lambda_{-2}(M_{2})$ (which fixes
the mass of the particle). This is a generalization of the 
result in the $\phi^4$ theory which is obtained by taking
$\lambda_{2}=0$ in Eq.\ (\ref{csexp2}). The standard derivation
in the $\phi^4$ theory of the range of values  of the 
coupling $\lambda_{0}$ for which perturbation theory is 
reliable~\cite{Willenbrock} can be improved by including 
the terms with $j\neq 0$ in Eq.\ (\ref{csexp2}). A measurement of
several cross sections, at high enough energy and with a
sufficient precision to be sensitive to the corrections due 
to the higher-dimensional terms, can be used to distinguish the 
expansion in Eq.\ (\ref{csexp2}) from the expansion of the most 
general effective Lagrangian and then to test the validity of
the reduction of couplings.

\subsection{Effective potencial of the effective scalar theory}

Another example of a systematic improvement of the $\phi^4$ theory 
analysis at the level of the effective theory is the application
of the renormalization group to the effective potential~\cite{Coleman}. 
The effective potential is the generating functional of one-particle 
irreducible Green functions evaluated at constant values of the field 
and its absolute minimum determines the ground state of the theory. 
It is an effective action (for constant fields) in the sense that it 
incorporates the effects of loop diagrams but it should not be confused 
with the action of the effective theory which incorporates the effects 
of the finite-energy range of validity of the theory. The effective 
potential of the effective theory incorporates both effects. 

The renormalization group equations for the one particle irreducible
Green functions of the effective theory yield a renormalization group 
equation for the effective potential ${\cal V}(\phi)$:

\begin{equation}
\mu \, {d \, {\cal V} \over d \mu} \, + \, {\vec \beta} \, 
{\partial \, {\cal V} \over \partial {\vec \lambda}} \, = \,
\gamma \, \phi \, {\partial \, {\cal V} \over \partial \phi}
\,, \label{RGEV}
\end{equation}

\noindent which is the generalization of the renormalization group
equation for the effective potential of the 
$\phi^4$ theory~\cite{Coleman} including all the parameters 
${\vec \lambda}$ of the effective theory and their corresponding 
$\beta$ functions already introduced in the discussion of the
renormalization of the effective theory in Sec.\ II. Following
the standard discussion~\cite{Coleman,Gross} of the renormalization 
group applied to the effective potential one introduces effective
couplings through the equations

\begin{equation}
{d {\vec {\bar \lambda}} \over d t} \, = \, 
{\vec \beta} ( {\vec {\bar \lambda}} )  \,,
\label{ec}
\end{equation}

\noindent with the initial conditions 
${\vec {\bar \lambda}}(0)={\vec \lambda}(\mu)$ and 
the rescaled field:

\begin{equation}
{\bar \phi} ( t ) \, = \, exp \left[ - \int_{0}^{t} dt' 
\gamma \left( {\vec {\bar \lambda}} ( t' ) \right) \right] \,
\phi \,.
\label{phi(t)}
\end{equation}

\noindent Then the renormalization group equation for the effective 
potential takes the simple form

\begin{equation}
{d \over d t} \, {\cal V} \left[ {\bar \phi} ( t )  , 
{\vec {\bar \lambda}} ( t )  , e^t \mu \right] \, = \, 0
\,, \label{RGEVt}
\end{equation}

\noindent which can be trivially solved leading to

\begin{equation}
{\cal V} \left[ \phi , {\vec \lambda} ( \mu ) , \mu \right]
\, = \, {\cal V} \left[ {\bar \phi} ( t )  , 
{\vec {\bar \lambda}} ( t )  , e^t \mu \right]
\,.\label{V(t)}
\end{equation}

Using simple dimensional analysis and obtaining the dependence
of the effective potential on the mass scale of the effective theory
$M_{2}$, one can use the renormalization group equation to determine
the behavior of the effective potential ${\cal V}$ as one scales the 
field $\phi$:

\begin{equation}
{\cal V} \left[ \phi , {\vec \lambda} ( \mu ) , M_{2} , \mu \right]
\, = \, e^{4t} \, {\cal V} \left[ e^{-t} {\bar \phi} ( t )  , 
{\vec {\bar \lambda}} ( t )  ,  e^{-t} M_{2}  ,  \mu \right]
\,. \label{V(t)2}
\end{equation}

\noindent The large logarithms which appear in a direct perturbative
calculation of the potential on the left-hand side dissappear on
the right-hand side if one makes the choice 
$t={1 \over 2}\ln (\phi^2 / \mu^2)$. This assumes
that the exponential factor depending on the anomalous dimension in 
Eq.\ (\ref{phi(t)}) is of order 1. For this choice of t one has 

\begin{equation}
{\vec {\bar \lambda}} ( t ) \, = \, {\vec \lambda} ( e^t \mu ) 
\, = \, {\vec \lambda} ( \phi )
\,. \label{l(phi)}
\end{equation}

The validity of the perturbative approach to the effective Lagrangian 
for $\mu\leq M_{2}$ then justifies a perturbative calculation of 
the effective potential for $\phi\leq M_{2}$. In lowest order (tree
level) one has 

\begin{equation}
{\cal V} \left[ \phi , {\vec \lambda} ( \mu ) , M_{2} , \mu \right]
\, = \, {\lambda_{-2} ( \phi ) M_{2}^2 \over 2} \, {\bar \phi}^2 ( t ) 
\, + \, {\lambda_{0} ( \phi ) \over 4!} \, {\bar \phi}^4 ( t ) \, + \,
{\lambda_{2}^{(1)} ( \phi ) \over 6! M_{2}^2} \, {\bar \phi}^6 ( t ) 
\, + \, ... \,,
\label{Vtree}
\end{equation}

\noindent i.e., the nonderivative terms in the effective Lagrangian with
the dimensionless parameters renormalized at a scale $\mu=
\phi$ and the field variable replaced by ${\bar \phi}(t)$.

The loop expansion for the effective potential in the $\phi^4$ theory
can be directly translated to the effective theory and then one has
a systematic approximation to the effective potential of the effective 
theory including radiative corrections. The consistency of the 
approach in Eq.\ (\ref{Vtree}), where corrections in inverse powers of the
mass scale of the effective theory are incorporated before considering
loop effects, requires 

\begin{equation}
{\lambda_{0} ( \phi ) \over 16 \pi^2}  \, \ll \,
{{\bar \phi}^2 \over M_{2}^2} \, \approx \, {\phi^2 \over M_{2}^2} 
\, \ll \, 1
\,. \label{cta}
\end{equation}

\noindent This is only possible if $\lambda_{0}(M_{2})
/ 16\pi^2 \ll 1$, i.e., if $M_{2}^2\ll M_{0}^2$ , and 
the approach will apply to the large field behavior of the
effective potential.

We end this discussion by pointing out that the consequences
in the effective potential of the reduction of couplings
at the level of the renormalization of the effective 
Lagrangian can be trivially identified since the effective 
potential is determined in terms of the renormalized 
parameters ${\vec \lambda}(\mu)$ of the effective Lagrangian.
This discussion also manifests the difference between 
the effective Lagrangian (an expansion in powers of the scalar
field and derivatives of the scalar field) and the nonanalytic
effective potencial.

\subsection{Spontaneous symmetry breaking and reduction of couplings}

In order to discuss a renormalized effective scalar field theory
with spontaneous symmetry breaking it is convenient to start by
rephrasing the mechanism of spontaneous symmetry breaking in the 
$\phi^4$ theory at the level of the renormalization group equations.
One considers a field $\eta$ which describes the fluctuations
around the vacuum of the theory and a Lagrangian

\begin{equation}
{\cal L}^{'} ( \eta ) \, = \, 
{1\over 2}  \partial_{\mu} \eta \partial^{\mu} \eta \,
- \, \lambda_{-3}^{'} M^3  \, \eta \, 
- \, {\lambda_{-2}^{'} M^2 \over 2} \, \eta^{2} \,
- \, {\lambda_{-1}^{'} M \over {3!}} \, \eta^{3} \,
- \, {\lambda_{0}^{'}\over {4!}} \, \eta^{4} \,
+ \, {\cal L}_{c.t.}^{'} ( \eta )
\,, \label{L'}
\end{equation}

\noindent which contains all the terms of dimension less or equal than 
four without any additional restriction and ${\cal L}_{c.t.}^{'}$ 
denotes the counterterms required to renormalize the theory. The 
coefficient of the linear term $\lambda_{-3}^{'}$ is fixed by the
condition that the vacuum expectation value of $\eta$ is zero
(in lowest order this leads to $\lambda_{-3}^{'}=0$ but this is
not so in higher orders). 

A one loop calculation leads to the renormalization group equations

\begin{equation}
\mu {d \lambda_{-3}^{'} \over d \mu} \, = \, - \,
\lambda_{-1}^{'} \, \lambda_{-2}^{'} \,,
\label{l'-3}
\end{equation}
\begin{equation}
\mu {d \lambda_{-2}^{'} \over d \mu} \, = \, 
{\lambda_{-1}^{'}}^2 \,
- \, \lambda_{0}^{'} \, \lambda_{-2}^{'} \,,
\label{l'-2}
\end{equation}
\begin{equation}
\mu {d \lambda_{-1}^{'} \over d \mu} \, = \, 3 \,
\lambda_{0}^{'} \, \lambda_{-1}^{'} \,,
\label{l'-1}
\end{equation}
\begin{equation}
\mu {d \lambda_{0}^{'} \over d \mu} \, = \, 3 \,
{\lambda_{0}^{'}}^2 \,,
\label{l'0}
\end{equation}

\noindent where an overall coefficient $1 / 16\pi^2$ has been
reabsorbed into a rescaling of all the dimensionless parameters. 

The case of spontaneous symmetry breaking corresponds to the 
possibility to find a translation of the variable
$\eta\to\phi =\eta +v$ such that when the
Lagrangian is written in terms of the variable $\phi$ only
terms invariant under the transformation $\phi\to -\phi$
appear,

\begin{equation}
{\cal L} ( \phi ) \, = \, {\cal L}^{'} ( \phi \, - \, v ) \, = \,
{1\over 2}  \partial_{\mu} \phi \partial^{\mu} \phi \,
- \, {\lambda_{-2} M^2 \over 2} \, \phi^{2} \,
- \, {\lambda_{0}\over {4!}} \, \phi^{4} \,
+ \, {\cal L}_{c.t.} ( \phi )
\,. \label{Lssb}
\end{equation}

The effect of a general translation of the field variable 
$\phi\to \phi + M v$ is equivalent
to a change of dimensionless parameters:

\begin{equation}
\lambda_{-3}^{'} \, = \, \lambda_{-3} \, + \,
v \, \lambda_{-2} \, + \, {1 \over 2} \, v^2 \, \lambda_{-1} \,
+ \, {1 \over 3!} \, v^3 \, \lambda_{0} \,,
\label{l'-3(l,v)}
\end{equation}
\begin{equation}
\lambda_{-2}^{'} \, = \, \lambda_{-2} \, + \,
v \, \lambda_{-1} \, + \, {1 \over 2} \, v^2 \, \lambda_{0} \,,
\label{l'-2(l,v)}
\end{equation}
\begin{equation}
\lambda_{-1}^{'} \, = \, \lambda_{-1} \, + \,
v \, \lambda_{0} \,, 
\label{l'-1(l,v)}
\end{equation}
\begin{equation}
\lambda_{0}^{'} \, = \, \lambda_{0} \,.
\label{l'0(l,v)}
\end{equation}

If one combines these relations, valid for an arbitrary 
translation of the field variables, with the equation determining
$\lambda_{-3}^{'}$ in terms of the remaining parameters in
${\cal L}^{'}$ ($\lambda_{-3}^{'}=0$ in lowest order) 
and uses the symmetry of ${\cal L}$ in the case of spontaneous symmetry breaking
($\lambda_{-3}=\lambda_{-1}=0$) then one can determine the
vacuum expectation value $v$ of the field $\phi$  in terms of the
parameters $\lambda_{-2}$ and $\lambda_{0}$. When the 
renormalization group equations (\ref{l'-3})--(\ref{l'0})
are rewritten in terms of the parameters $\lambda_{k}$
corresponding to this value of $v$ then one finds

\begin{equation}
\mu {d \lambda_{-3} \over d \mu} \, = \,
\mu {d \lambda_{-1} \over d \mu} \, = \, 0 \,,
\label{l-3,-1}
\end{equation}

\noindent which is the manifestation at this level of the 
renormalizability of the $\phi^4$ theory with spontaneous symmetry 
breaking, and the renormalization group equations for the two 
parameters of the theory become

\begin{eqnarray}
\mu {d \lambda_{-2} \over d \mu} \,& = &\, 5 \, \lambda_{0} \,
\lambda_{-2} \\
\mu {d \lambda_{0} \over d \mu} \,& = &\, 3 \, \lambda_{0}^2
\,.\label{RGESSB}
\end{eqnarray}

Once the spontaneous breaking of the symmetry has been 
formulated at the level of the renormalization group equations
for the $\phi^4$ theory in a mass-independent renormalization 
scheme the generalization to the case of the effective scalar
theory is trivial. Instead of the Lagrangian ${\cal L}^{'}$ in Eq.\
(\ref{L'}) one has to consider all possible terms with any
dimension and instead of the renormalization group equations
in Eqs.\ (\ref{l'-3})--(\ref{l'0}) one now has the renormalization 
group equations for the parameters of the effective scalar
theory, including the 
parameters ${\vec \lambda}_{2n+1}^{'}$ associated to terms
with an odd number of fields. The structure of the renormalization
group equations, based on dimensional arguments and the 
expansion in powers of $\lambda_{0}$ , is not modified by
the addition of odd terms. Once more a translation $v$ of the
field variable is determined such that the effective Lagrangian
in the translated field variable is invariant under $\phi\to
-\phi$ . Its value is determined as a double expansion
in powers of $\lambda_{0}$ and $\lambda_{-2}$ from the condition
that the vacuum expectation value of the field $\eta$ vanishes.

When the renormalization group equations for the parameters
${\vec \lambda}^{'}$ of the effective theory are written in
terms of the parameters of the Lagrangian which results
from a translation of the field variable by the 
value of $v$ determined previously then one finds once more
Eq.\ (\ref{l-3,-1}) and 

\begin{equation}
\mu {d {\vec \lambda}_{2n+1} \over d \mu} \, = \, 0
\label{l2n+1}
\end{equation}

\noindent which is the manifestation at this level of the 
renormalizability of the effective theory with spontaneous symmetry 
breaking. One also has renormalization group equations for the
parameters $\lambda_{-2}$, $\lambda_{0}$, $\lambda_{2n}^{i_n}$
which are the analogue for the effective scalar theory with
spontaneous symmetry breaking of the renormalization group equations 
of the effective theory in the symmetric case discussed in Sec.\ III.

Since the structure of the renormalization group equation for 
$\lambda_{2n}^{i_n}$ is fixed by dimensional arguments which are 
not affected by the spontaneous breaking of the symmetry, the 
step-by-step determination of a reduction couplings applies also
in this case. In fact the lowest-order term in the expansion in
powers of $\lambda_{-2}$ (zero-order term), which is the term
identifying the possible reductions of couplings, is the one found 
in the symmetric case. Then one has a one-to-one correspondence 
between the reduction of couplings in the symmetric and 
spontaneously broken cases and the differences appear in the 
extension of the reduction to higher orders in the expansion in 
powers of $\lambda_{-2}$.

The conclusion of this discussion is that there is no obstruction 
to translating all the analysis of the symmetric effective theory
to the case of spontaneous symmetry breaking. This means that 
all the bounds found in the $\phi^4$ theory~\cite{Dashen} with
spontaneous symmetry breaking can be
discussed also at the level of the effective theory as we argued
before in the symmetric case. In particular the stability bounds 
on the theory derived from the large field behavior
of the effective potential for a scalar field coupled to 
fermionic fields~\cite{Cabibbo} can be an example where
the higher-dimensional terms in the effective theory can be 
important depending on the ratio of scales of the effective 
theory. 

\section{Summary and Outlook}

It has been shown that the perturbative renormalization of the 
theory of a scalar field with nonrenormalizable couplings
(negative dimension) which requires an infinite number of
counterterms, is compatible with the presence of only a finite
number of independent parameters. The renormalization group 
equations of the effective theory allow us to identify three
possible extensions of the $\phi^4$ theory with an additional
dimensionless parameter $\lambda_{2}$ which controls the contribution 
of all the higher-dimensional terms. The extension is determined order 
by order in a double expansion in the quartic coupling $\lambda_{0}$
and the product $\lambda_{-2}\lambda_{2}$, where $\lambda_{-2}$ is the
dimensionless parameter associated to the mass term.

Assuming the validity of perturbation theory and of the 
effective theory energy expansion, it is possible to improve
systematically the bounds on the $\phi^4$ theory due to  
triviality both in the symmetric and the spontaneously broken cases. 
If one assumes that the fundamental theory at higher
energies is such that its low-energy limit is described by one of the 
minimal extensions of the $\phi^4$ theory then the modifications to the 
bounds due to the higher dimensional terms can be calculated in terms 
of just one additional parameter with respect to the renormalizable 
$\phi^4$ theory. An analysis along these lines of a minimal extension of
the standard model could have important physical implications if the
characteristic mass scale of this effective theory is
not much higher than the Fermi scale. The first step in this direction would be
to generalize the discussion of the effective scalar field theory
including a set of fermionic fields coupled to a 
set of scalar fields  corresponding to the standard model neglecting 
the gauge interactions.

Another possible extension of the present discussion of the method
of reduction of couplings is to consider the possibility of a ressumation 
of terms in the double expansion which goes beyond the step-by-step
reduction of couplings considered in this work. The possibility to have 
a  fixed point of the renormalization group equations of an effective 
theory not corresponding to a free theory and having a finite-dimensional
domain of attraction (asymptotic safety~\cite{WeinbergIII}) can be
seen as an example of a reduction of couplings. The restriction to the
finite-dimensional surface, consisting of the trajectories of the 
renormalization group which are attracted into the fixed point, defines
a set of infinite relations among couplings compatible with 
renormalizability leading to a theory with a finite number of free 
parameters. 

The interpretation of a lattice $\phi^4$ theory as an effective scalar
field theory, based on the local effective Lagrangian which reproduces
the asymptotic small lattice spacing expansion of Green functions, 
provides a new framework with which to discuss 
the reduction of couplings. In fact
the improvement program~\cite{Symanzik} can be interpreted as the 
identification of suitable irrelevant terms to be added to the
lattice action leading to an effective scalar theory whose first
correction to the renormalizable $\phi^4$ theory has higher and 
higher dimension as one goes to higher and higher orders in the 
improvement. The final result of the improvement program
is a lattice field theory which is an effective field theory with
a trivial reduction of couplings (all the parameters corresponding
to irrelevant terms in the effective Lagrangian vanish). Each step
in the improvement can be seen as a reduction of couplings of a 
different type of the reduction of couplings considered in this work.
The improvement at order k corresponds to $\lambda_{2n}^{i_n} = 0$,
$0 < n < k$, which is another way to fix arbitrary parameters in the
effective Lagrangian in a way compatible with the renormalization 
group equations. On the other hand, the reduction of couplings we
have studied corresponds to a different way to fix arbitrary parameters;
one first looks for renormalization group invariant relations among
the parameters $\lambda_{2}^{i_1}$, next one finds expressions for 
$\lambda_{4}^{i_2}$ in terms of the independent parameter of the
previous step, and one extends the procedure order by order in the
energy expansion of the effective Lagrangian.

The identification of a lattice action, with irrelevant terms fixed 
in such a way that its local effective Lagrangian coincides with the
effective Lagrangian of one of the minimal extensions of the 
renormalizable $\phi^4$ theory, suggests the possibility of a 
reformulation of the reduction of couplings at the level
of lattice field theory.
This could allow the discussion of the method of reduction of couplings 
at the nonperturbative level.
Finally, it would be interesting to find an example of the realization
of the idea of reduction of couplings in a simple enough theory to be
able to determine its low-energy limit as a minimal extension of a
renormalizable theory.

\bigskip

\bigskip\leftline{{\bf Acknowledgments}}\medskip
This work was partially supported by CICYT contract AEN 96-1670.  
The work of M.A. has been supported by DGA.
 
\vfill
\eject

\newpage

\end{document}